\documentclass[aps,pra,twocolumn,superscriptaddress,showpacs]{revtex4}
\def\be{\begin{equation}}
\def\ee{\end{equation}}
\def\bea{\begin{eqnarray}}
\def\eea{\end{eqnarray}}
\def\bi{\begin{itemize}}
\def\ei{\end{itemize}}

\usepackage{graphicx}
\usepackage{amsmath}
\usepackage{amssymb}

\begin{document}

\title{ 
Adiabatic dynamics of an inhomogeneous quantum phase transition:\\
             the case of $z>1$ dynamical exponent 
}

\author{Jacek Dziarmaga}

\affiliation{ Institute of Physics and 
              Centre for Complex Systems Research, 
              Jagiellonian University,
              Reymonta 4, 30-059 Krak\'ow, 
              Poland}

\author{Marek M. Rams}

\affiliation{ Institute of Physics and 
              Centre for Complex Systems Research, 
              Jagiellonian University,
              Reymonta 4, 30-059 Krak\'ow, 
              Poland}

\affiliation{ Theoretical Division, Los Alamos National Laboratory, MS-B213, 
              Los Alamos, NM 87545, USA}

\date{May 18, 2010}

\begin{abstract}
We consider an inhomogeneous quantum phase transition across a multicritical point 
of the XY quantum spin chain. This is an example of a Lifshitz transition with 
a dynamical exponent $z=2$. Just like in the case $z=1$ considered in New J. Phys. 
{\bf 12}, 055007 (2010) when a critical front propagates much faster than the maximal 
group velocity of quasiparticles $v_q$, then the transition is effectively homogeneous: 
density of excitations obeys a generalized Kibble-Zurek mechanism and scales with the 
sixth root of the transition rate. However, unlike for $z=1$, the inhomogeneous transition 
becomes adiabatic not below $v_q$ but a lower threshold velocity $\hat v$, proportional 
to inhomogeneity of the transition, where the excitations are suppressed exponentially. 
Interestingly, the adiabatic threshold $\hat v$ is nonzero despite vanishing minimal group 
velocity of low energy quasiparticles. In the adiabatic regime below $\hat v$ the inhomogeneous 
transition can be used for efficient adiabatic quantum state preparation in a quantum simulator: 
the time required for the critical front to sweep across a chain of $N$ spins adiabatically is 
merely linear in $N$, while the corresponding time for a homogeneous transition across the 
multicritical point scales with the sixth power of $N$. What is more, excitations after the 
adiabatic inhomogeneous transition, if any, are brushed away by the critical front to the end 
of the spin chain. 
\end{abstract}

\maketitle

\section{ INTRODUCTION }
\label{Intro}

A quantum phase transition is a qualitative change in the ground state of 
a quantum system when one of the parameters in its Hamiltonian passes through
a critical point. In a second order transition a continuous change is 
accompanied by a diverging correlation length and vanishing energy gap. The 
vanishing gap implies that no matter how slowly a system is driven 
through the transition its evolution cannot remain adiabatic near the 
critical point. As a result, after the transition the system is excited 
to a state with a finite correlation length $\hat\xi_{KZ}$ whose size shrinks 
with increasing rate of the transition. This scenario, known as Kibble-Zurek (KZ)
mechanism (KZM), was first described in the context of finite temperature 
transitions \cite{K,Z}. Although originally motivated by cosmology \cite{K}, 
KZM at finite temperature was confirmed by numerical simulations of 
the time-dependent Ginzburg-Landau model \cite{KZnum} and successfully tested by 
experiments in liquid crystals \cite{LC}, superfluid helium 3 \cite{He3}, both 
high-$T_c$ \cite{highTc} and low-$T_c$ \cite{lowTc} superconductors, and 
convection cells \cite{ne}. More recently, spontaneous appearance of vortexes
during Bose-Einstein condensation driven by evaporative cooling was observed in 
Ref. \cite{Anderson}. However, the quantum zero temperature limit, which is in 
many respects qualitatively different, remained unexplored until recently, see
e.g. Refs. \cite{3sites,Bodzio1,KZIsing,Dziarmaga2005,Polkovnikov,Levitov,JDrandom,
Cucchietti,Bodzioferro,Hindusi,others} and Ref. \cite{JDreview} for a review. The recent 
interest is motivated in part by adiabatic quantum computation or adiabatic quantum state 
preparation, where one would like to cross a quantum critical point as adiabatically as 
possible, and in part by condensed matter physics of ultracold atoms, where it is easy 
to manipulate parameters of a Hamiltonian in time and which, unlike their solid state 
physics counterparts, are fairly well isolated from their environment. In fact, an 
instantaneous quench to the ferromagnetic phase in a spinor BEC resulted in finite-size 
ferromagnetic domains whose origin was attributed to KZM \cite{ferro}. However, since the 
transition rate was formally infinite in that experiment, the KZ scaling relation between 
the average domain size $\hat\xi_{KZ}$ and the quench rate has not been verified. 

The KZM argument is briefly as follows \cite{Z,KZIsing}. When a transition is driven 
by varying a parameter $g$ in the Hamiltonian across an isolated critical point at 
$g_c$, then we can define a dimensionless distance from the critical point as
\be
\epsilon~=~\frac{g-g_c}{g_c}~.
\ee
When $\epsilon\to0$ the correlation length $\xi$ in the ground state diverges as 
$\xi\sim|\epsilon|^{-\nu}$, and the energy gap $\Delta$ between the ground state
and the first excited state vanishes as $\Delta\sim|\epsilon|^{z\nu}$. 
Setting $\hbar=1$ from now on, a diverging $\Delta^{-1}\sim|\epsilon|^{-z\nu}$ is the 
shortest time scale on which the ground state can adjust adiabatically to varying 
$\epsilon(t)$. A generic $\epsilon(t)$ can be linearized near the critical point
$\epsilon=0$ as
\be
\epsilon(t)~\approx~-\frac{t}{\tau_Q}~+~{\cal O}(t^2),
\label{tauQ}
\ee
where the coefficient $\tau_Q$ is called a quench time. Assuming that the system
was initially prepared in its ground state, its adiabatic evolution fails at an 
$\hat\epsilon_{KZ}$ when time $\hat t_{KZ}$ left to crossing the critical point equals 
the shortest time scale $\Delta^{-1}$ on which the ground state can adjust. Solving 
this equality, we obtain
\bea
\hat\epsilon_{KZ} &\sim& \tau_Q^{-\frac{1}{z\nu+1}}~,   \label{hatepsilon}\\ 
\hat t_{KZ}       &\sim& \tau_Q^{\frac{z\nu}{z\nu+1}}~. \label{hatt}      
\eea
From $\hat\epsilon_{KZ}$ the evolution becomes impulse, i.e. the state does not evolve
but remains frozen in the ground state at $\hat\epsilon_{KZ}$, until 
$-\hat\epsilon_{KZ}$ when the evolution becomes adiabatic again. In this way, 
the ground state at $\hat\epsilon_{KZ}$ with a KZ correlation length
\be
\hat\xi_{KZ} ~\sim~ \hat\epsilon_{KZ}^{-\nu} ~\sim~ \tau_Q^{\frac{\nu}{z\nu+1}}
\label{hatxi}
\ee
becomes the initial excited state for the adiabatic evolution after $-\hat\epsilon_{KZ}$. 
This $\hat\xi_{KZ}$ determines density of quasiparticles excited during a phase transition in, 
say, one dimension
\be
d ~\sim~ \hat\xi_{KZ}^{-1} ~\sim~ \tau_Q^{-\frac{\nu}{z\nu+1}}~
\label{dKZgeneral}
\ee
and, in general, expectation values of other operators according to their critical scaling 
dimensions. An operator $O$ whose expectation value scales like $\langle O\rangle\sim\xi^\phi$
in the ground state near the critical point will scale like $\langle O\rangle\sim\hat\xi_{KZ}^\phi$ 
right after the dynamical transition.

Note that when $\tau_Q$ is large enough, then $\hat\epsilon_{KZ}$ is small and the linearization 
in Eq. (\ref{tauQ}) is self-consistent: the KZM physics happens very close to the critical 
point between $-\hat\epsilon_{KZ}$ and $+\hat\epsilon_{KZ}$.

Recently the quantum KZ paradigm has been generalized to transitions that happen in space rather than in time
\cite{Karevski,Dorner,Damski,inhomRams} and to inhomogeneous transitions that happen in time but are
inhomogeneous in space \cite{inhomRams,Schaller}. They are described in more detail in the following
Sections \ref{inhom_general} and \ref{KZMinspace_general} respectively. In an inhomogeneous transition
it is important how fast the critical front, i.e., the place where the control parameter $\epsilon$ 
is zero moves in space. In Ref. \cite{inhomRams} we considered an example of the quantum Ising chain which
is a representative of a universality class with the dynamical exponent $z=1$. This exponent means that
at the critical point the dispersion of low energy quasiparticles is linear with a unique quasiparticle
velocity $v_q$. Reference \cite{inhomRams} demonstrates that when a critical front is moving with a velocity 
$v\gg v_q$, then the transition proceeds as it were effectively homogeneous and the homogeneous estimate 
(\ref{dKZgeneral}) applies. On the other hand, when $v\ll v_q$ then the excitations are exponentially 
suppressed and the transition is effectively adiabatic. 

This picture becomes a bit more complex when $z>1$ like in the Lifshitz transition with $z=2$. When $z>1$ 
there is a non-linear low energy quasiparticle dispersion at the critical point $\omega\sim k^z$ and no unique 
quasiparticle velocity. However, similar casuality arguments as for $z=1$ lead to a more general 
conclusion that the transition is effectively homogeneous when $v\gg v_q$ with $v_q$ being the maximal 
group velocity of quasiparticles at the critical point (equal to the unique quasiparticle velocity when 
$z=1$). Indeed, when the critical front is much faster than $v_q$, than even the fastest quasiparticles 
are not able to communicate any information across the front and the transition is effectively homogeneous. 
However, we also demonstrate that the condition $v\ll v_q$ is not enough for an inhomogeneous transition to 
become adiabatic when $z>1$, but at the same time one does not need to go as far as below the minimal 
quasiparticle group velocity which is zero when $z>1$. The inhomogeneous transition turns out to be
adiabatic below a finite threshold velocity $\hat v$, proportional to an inhomogeneity of the transition, 
which can be obtained from a variation of the simple KZ argument. In this paper we both present the general 
physical arguments and support them by a solution of a transition across a multicritical point of the XY model 
\cite{multicritical}. The solution demonstrates that the general arguments are robust despite their simplicity, 
while the general discussion suggests that the conclusions are applicable beyond this particular example. 

The paper is organized as follows. In the following Section \ref{inhom_general} we provide experimental 
motivation and a generalized KZ argument for an inhomogeneous transition. It is here that we obtain the 
general estimate for $\hat v$. In Section \ref{KZMinspace_general} we do the same for a static transition 
in space and then we use the general results for a static transition to derive again the same estimate 
for $\hat v$ as in Section \ref{inhom_general}. In this way the general $\hat v$ is arrived at from two 
different angles. In order to test the general predictions we introduce the transverse field XY model in 
Section \ref{SectionXY}. In Section \ref{XYhomo} we review an exact solution for a homogeneous transition 
in this model \cite{multicritical}, and in Subsection \ref{XYhomoadiab} a homogenous transition in a finite 
chain of $N$ spins. The last transition becomes adiabatic when the transition time $\tau_Q$ is much longer 
than $N^6$. In Section \ref{XYinspace} we consider a transition in space and confirm the general predictions 
of Section \ref{KZMinspace_general} in the XY model. Section \ref{XYinhomo} is devoted to an inhomogeneous 
transition in the XY model. In its Subsection \ref{XYinhomohomo} the effectively homogeneous case of $v\gg v_q$ 
is studied, and in Subsections \ref{XYinhomoadiab} and \ref{XYinhomoedge} the adiabatic regime of $v\ll\hat v$.   
In Subsection \ref{XYinhomoadiab} $\hat v$ is shown to be proportional to the inhomogeneity of the transition 
at the critical front. It is argued there that the adiabatic quantum state preparation by an inhomogeneous 
transition requires time proportional to the number of spins $N$, i.e., for large $N$ the inhomogeneous transition 
is much faster than the adiabatic homogeneous transition. This conclusion is further strenghened in Subsection 
\ref{XYinhomoedge} showing that eventual (exponentially small) quasiparticle excitations in the adiabatic 
inhomogeneous transition are brushed away by the critical front to the end of the spin chain. Finally, the 
concluding Section \ref{Conclusion} provides a summary of our results.

\section{ Inhomogeneous transition }
\label{inhom_general}

As pointed out already in the finite temperature context \cite{Volovik}, see
also Ref. \cite{VolovikRecent} for recent applications, in a realistic 
experiment it is difficult to make $\epsilon$ exactly homogeneous throughout a 
system. For instance, in the superfluid $^3He$ experiments \cite{He3} the
transition was caused by neutron irradiation of helium 3. Heat released in each
fusion event, $n~+~^3He~\to~^4He$, created a bubble of normal fluid above the 
superfluid critical temperature $T_c$. As a result of quasiparticle diffusion, 
the bubble was expanding and cooling with local temperature 
$T(t,r)=\exp(-r^2/2Dt)/(2\pi Dt)^{3/2}$, where $r$ is a distance from the center 
of the bubble and $D$ is a diffusion coefficient. Since this $T(t,r)$ is hottest 
in the center, the transition back to the superfluid phase, driven by an 
inhomogeneous parameter 
\be
\epsilon(t,r)~=~\frac{T(t,r)-T_c}{T_c}~,
\ee 
proceeded from the outer to the central part of the bubble with a critical front 
$r_c(t)$, where $\epsilon=0$, shrinking with a finite velocity $v=dr_c/dt<0$.

A similar scenario is generic in ultracold atom gases in magnetic/optical 
traps \cite{Greiner,Scalettar}: a trapping potential results in inhomogeneous 
density of atoms $\rho(\vec r)$ and a critical point $g_c$ depends 
on atomic density $\rho$. Thus even a transition driven by a perfectly homogeneous 
$g(t)$ is inhomogeneous,
\be
\epsilon(t,\vec r)~=~
\frac{g(t)-g_c[\rho(\vec r)]}{g_c[\rho(\vec r)]}~,      
\ee
with a surface of critical front, where $\epsilon=0$, moving at a finite speed. 

According to the KZM, in a homogeneous symmetry breaking transition, a state after the 
transition is a mosaic of finite ordered domains of average size $\hat\xi_{KZ}$. Within 
each finite domain orientation of the order parameter is constant but uncorrelated to
orientations in other domains. In contrast, in an inhomogeneous symmetry breaking 
transition \cite{Volovik}, the parts of the system that cross the critical point 
earlier may be able to communicate their choice of orientation of the order parameter
to the parts that cross the transition later and bias them to make the same choice. 
Consequently, the final state may be correlated at a range longer than $\hat\xi_{KZ}$ or 
even end up being the ground state, and the final density of excited quasiparticles 
may be lower than the KZ estimate in Eq. (\ref{dKZgeneral}) or even zero. 

From the point of view of testing KZM, this inhomogeneous scenario, when relevant, may 
sound like a negative result because an imperfect inhomogeneous transition suppresses KZM.
However, from the point of view of adiabatic quantum computation or adiabatic quantum state 
preparation it is the KZM itself that is a negative result: no matter how slow the 
homogeneous transition is there is a finite density of excitations (\ref{dKZgeneral}) 
which decays only as a fractional power of transition time $\tau_Q$. From this perspective, 
the inhomogeneous transition may be a practical way to suppress KZ excitations and prepare 
the desired final ground state adiabatically.

To estimate when the inhomogeneity may actually be relevant, in a similar way as in 
Eq. (\ref{tauQ}) and Ref. \cite{inhomRams}, we linearize the parameter $\epsilon(t,n)$ 
in both $n$ and $t$ near the critical front where $\epsilon(t,n)=0$:
\be
\epsilon(t,n)~\approx~\alpha~(n-vt)~.
\label{alpha}
\ee
Here $n$ is position in space, e.g. lattice site number, $\alpha$ is a gradient/inhomogeneity 
of the transition, and $v$ is velocity of the critical front. When watched locally at 
a fixed $n$, the inhomogeneous transition in Eq. (\ref{alpha}) appears to be the homogeneous 
transition in Eq. (\ref{tauQ}) with a local
\be
\tau_Q~=~\frac{1}{\alpha v}~.
\label{tauQalphav}
\ee    
The part of the system where $n<vt$, or equivalently $\epsilon(t,n)<0$, is already
in the broken symmetry phase. An outcome of the transition depends on $v$.

On one hand, there cannot be efficient communication across the critical point when it 
is moving faster than quasiparticles near the critical point:
\be
v~\gg~v_q~.
\label{vq}
\ee
Here $v_q$ is the maximal group velocity of quasiparticles at $\epsilon=0$ or, in general, 
a Lieb-Robinson velocity \cite{LiebRobinson}. It is a constant that does not depend on the 
inhomogeneity $\alpha$. In this ``homogeneous regime'' the inhomogeneous transition is effectively 
homogeneous and the final density of excitations after the transition is given by 
Eq. (\ref{dKZgeneral}) with the local $\tau_Q=1/\alpha v$. 

On the other hand, KZM provides the relevant scales of length and time, $\hat\xi_{KZ}$ and $\hat t_{KZ}$ 
respectively, whose combination \cite{Volovik}
\be
\hat v ~\simeq~ \frac{\hat\xi_{KZ}}{\hat t_{KZ}} ~ \simeq ~\alpha^{\frac{\nu(z-1)}{1+\nu}}   ~.
\label{hatv}
\ee 
is a relevant scale of velocity. Here we used Eqs. (\ref{hatt},\ref{hatxi}) and 
Eq. (\ref{tauQalphav}) which is valid for small $\alpha$. Indeed, when $v\ll\hat v$ the system has 
enough time to adjust adiabatically on the relevant length $\hat\xi_{KZ}$ and the density of excitations 
is less than in the homogeneous KZM. This is an ``adiabatic regime'' of the inhomogeneous transition.

In general the adiabatic threshold $\hat v$ depends on the inhomogeneity $\alpha$
in distinction to the constant $v_q$. However, in a special case of $z=1$, or 
linear quasiparticle dispersion at the critical point, the two velocities are the 
same: $\hat v\simeq v_q$. This (quite generic) special case was studied recently in 
Refs. \cite{inhomRams,Schaller}. In this paper we consider an example of a more general 
class of critical points with $z>1$, or nonlinear quasiparticle dispersion, when the adiabatic 
threshold $\hat v$ in Eq. (\ref{hatv}) scales with a positive power of $\alpha$ and,
for small enough $\alpha$, can be clearly distinguished from the homogeneous threshold $v_q$: 
$\hat v\ll v_q$.

However, before we proceed with the example, in the next Section we rederive 
Eq. (\ref{hatv}) from a slightly different perspective.

\section{ KZM in space }
\label{KZMinspace_general}

References \cite{Dorner,Damski,inhomRams,Karevski} considered a ``phase transition 
in space'' where $\epsilon(n)$ is inhomogeneous but time-independent. In the same 
way as in Eq. (\ref{alpha}), this parameter can be linearized as
\be
\epsilon(n)~\approx~\alpha~(n-n_c)~,
\label{nc}
\ee
near the static critical front at $n=n_c$ where $\epsilon=0$. The system is in 
the broken symmetry phase where $n<n_c$ and in the symmetric phase where $n>n_c$. 
In the first ``local approximation'', we would expect that the order parameter behaves 
as if the system were locally  homogeneous: it is nonzero for $n<n_c$ only, and when 
$n\to n_c^-$ it tends to zero as $(n_c-n)^\beta$ with the critical exponent $\beta$. 
However, this first approximation is in contradiction 
with the basic fact that the correlation (or healing) length $\xi$ diverges as
$\xi\sim|\epsilon|^{-\nu}$ near the critical point and the diverging $\xi$ is 
the shortest length scale on which the order parameter can adjust to (or heal 
with) the changing $\epsilon(n)$. Consequently, when approaching $n_c^-$ the local
approximation $(n_c-n)^\beta$ must break down when the local correlation length
$\xi\sim[\alpha(n_c-n)]^{-\nu}$ equals the distance remaining to the critical point
$(n_c-n)$. Solving this equality with respect to $\xi$, we obtain 
\be
\hat\xi_{SP}~\sim~\alpha^{-\frac{\nu}{1+\nu}}~.
\label{tildexigeneral}
\ee 
Beginning from $n-n_c\simeq-\hat\xi_{SP}$ the ``evolution'' of the order parameter in $n$ 
becomes ``impulse'', i.e, the order parameter does not change until $n-n_c\simeq+\hat\xi_{SP}$ 
in the symmetric phase, where it begins to follow the local $\epsilon(n)$ again and 
decays to zero on the same length scale of $\hat\xi_{SP}$. 

A direct consequence of this ``KZM in space'' is that a non-zero order parameter 
penetrates into the symmetric phase to a depth
\be
\delta n ~\sim~ \hat\xi_{SP} ~
\label{deltangeneral}
\ee 
as if the the critical point were effectively ``rounded off'' on the length scale of 
$\hat\xi_{SP}$. This rounding-off results also in a finite energy gap 
\be
\hat\Delta_{SP}~\sim~\hat\xi_{SP}^{-z}~\sim~\alpha^{\frac{z\nu}{1+\nu}}~
\label{tildeDeltageneral}
\ee
in contrast to the local approximation, where we would expect gapless excitations 
near the critical point. The finite gap in turn should prevent excitation of the 
system even when the critical point $n_c$ in Eq. (\ref{nc}) moves with a finite 
velocity: $n_c(t)=vt$. The excitation is suppressed up to a threshold velocity
\be
\hat v~\sim~
\frac{\hat\xi_{SP}}{\hat\Delta_{SP}^{-1}}~\sim~
\alpha^{\frac{\nu(z-1)}{1+\nu}}
\label{tildevgeneral}
\ee
which is a ratio of the relevant length $\hat\xi_{SP}$ to the relevant time $\hat\Delta_{SP}^{-1}$. 
This $\hat v$ is the same as Eq. (\ref{hatv}). 

In the following Sections we test the general predictions in the XY model.

\section{ Transverse field XY chain } 
\label{SectionXY}

The ferromagnetic transverse field XY quantum spin chain is
\bea
H=
-\sum_{n=1}^N
h_n \sigma^z_n
-\sum_{n=1}^{N-1}
\left(
J^x_n\sigma^x_n\sigma^x_{n+1}+
J^y_n\sigma^y_n\sigma^y_{n+1}
\right),
\label{Hsigma}
\eea 
where $\sigma^{x,y,z}$ are spin-$1/2$ Pauli matrices. Here we consider a path in the 
parameter space 
$(J^x,J^y,h)=\left(\frac{1-\epsilon}{2},\frac{1+\epsilon}{2},1+\epsilon\right)$ 
parametrized by the anisotropy of the ferromagnetic couplings $\epsilon\in[-1,1]$.
The parameter $\epsilon$ will be driven from the initial $\epsilon=1$ to the final
$\epsilon=-1$, when the Hamiltonian (\ref{Hsigma}) becomes the simple Ising chain
\be
H_{\rm final}~=~-\sum_{n=1}^{N-1}\sigma^x_n\sigma^x_{n+1}~.
\label{Hfinal}
\ee
In the thermodynamic limit a homogeneous system has a second order quantum phase 
transition (multicritical point) at $\epsilon=0$, which separates a paramagnetic 
phase where $\epsilon>0$ from a ferromagnetic phase where $\epsilon<0$. Non-adiabaticity
of this dynamical transition will be quantified by average number of 
quasiparticle/kink excitations in the final ferromagnetic state.

In a more general inhomogeneous system that we will consider in this paper it is 
natural to parametrize
\bea
J^x_n~=~\frac{1-\epsilon_{n+\frac12}}{2}~,~~
J^y_n~=~\frac{1+\epsilon_{n+\frac12}}{2}~,~~
h_n~=~1+\epsilon_n~,
\nonumber
\eea
where $\epsilon_n(t)$ is a continuous function of $n$ and, in general, time.

After the Jordan-Wigner transformation to spinless fermionic annihilation operators $c_n$,
$
\sigma^x_n=\left( c_n - c_n^\dagger \right) i S_n
$,
$
\sigma^y_n=\left( c_n + c_n^\dagger \right)S_n
$,
$
\sigma^z_n=1-2 c^\dagger_n c_n
$,
where $S_n=\prod_{m<n}(1-2 c^\dagger_m c_m)$, the Hamiltonian (\ref{Hsigma}) becomes  
\bea
H &=& 
\sum_n 
\left[
-\left(c_n^\dag c_{n+1}+{\rm h.c.}\right)-
\epsilon_{n+\frac12}
\left(c_{n+1}c_n+{\rm h.c.}\right)
\right.
\nonumber\\
& &
\left.
+\left(1+\epsilon_n\right)
\left(2c^\dag_n c_n-1\right)~ 
\right] ~.
\label{Hc}
\eea
This quadratic $H$ is diagonalized to $H=\sum_m\omega_m\gamma_m^\dag\gamma_m$ by 
a Bogoliubov transformation 
$c_n=\sum_{m=0}^{N-1}(u_{nm}\gamma_m + v^*_{nm} \gamma_m^\dagger)$ with $m$ numerating 
$N$ eigenmodes of the stationary Bogoliubov-de Gennes equations,
\bea
\omega u^\pm_n &=&
2(1+\epsilon_n) u^\mp_n 
\nonumber\\
&-&
(1\mp\epsilon_{n+\frac12}) u^\mp_{n+1}-(1\pm\epsilon_{n+\frac12}) u^\mp_{n-1}   
~,
\label{BdG}
\eea
with $\omega\geq0$. Here $u^\pm\equiv u \pm v$. 

The Hamiltonian (\ref{Hsigma},\ref{Hc}) commutes with a parity operator
\be
P~=~\prod_{n=1}^N\sigma^z_n~=~\prod_{n=1}^N\left(2c_n^\dag c_n-1\right)~.
\ee
The even parity of the initial ground state at $\epsilon=1$ is conserved during 
the dynamical transition.

\section{ Homogeneous transition } 
\label{XYhomo}

When $N\to\infty$ and $\epsilon_n=\epsilon$ is homogeneous, it is convenient to 
make a Fourier transform 
$u^\pm_k=N^{-1/2}\sum_{n=1}^Nu^\pm_ne^{-ikn}$ with pseudomomentum 
$k\in(-\pi,\pi]$. Equation (\ref{BdG}) becomes
\be
\frac{\omega_k}{2}
\left(
\begin{array}{c}
u^+_k \\
u^-_k
\end{array}
\right)=
\left[\sigma^x(1+\epsilon-\cos k)-\sigma^y\epsilon\sin k\right]
\left(
\begin{array}{c}
u^+_k \\
u^-_k
\end{array}
\right)~,
\label{BdGk}
\ee
with eigenfrequencies
\be
\omega_k=2\sqrt{(1+\epsilon-\cos k)^2+\epsilon^2\sin^2 k}
\label{omegak}
\ee
We can expand $\omega_0\approx2|\epsilon|\equiv|\epsilon|^{z\nu}$ for small $\epsilon$, 
and $\omega_k\approx k^2\equiv|k|^z$ for $\epsilon=0$ and small $k$, to identify the 
critical exponents:
\be
z~=~2~,~~\nu=1/2~.
\label{znu}
\ee 
With these exponents, the general Eq. (\ref{dKZgeneral}), valid for an isolated quantum 
critical point, would predict the density of excitations after the homogeneous transition 
in Eq. (\ref{tauQ}) to scale as $d\simeq\tau_Q^{-1/4}$. However, as shown in Ref. 
\cite{multicritical} and outlined below, the multicritical nature of the critical point 
results in a different scaling exponent and the KZM requires careful generalization 
here \cite{multicritical,JDreview}. 

The correct scaling can be obtained by mapping our time-dependent problem 
\be
\frac{i}{2}\frac{d}{dt}
\left(
\begin{array}{c}
u^+_k \\
u^-_k
\end{array}
\right)=
\left[
\sigma^x
\left(1-\frac{t}{\tau_Q}-\cos k\right)+
\sigma^y\frac{t}{\tau_Q}\sin k
\right]
\left(
\begin{array}{c}
u^+_k \\
u^-_k
\end{array}
\right)~,
\label{tBdGk}
\ee
to the standard Landau-Zener (LZ) model \cite{LZ}:
\be
i\frac{d}{dt'}
\left(
\begin{array}{c}
u^+_k \\
u^-_k
\end{array}
\right)=
\left[
~\frac{t'}{\tau}~\sigma~+~\Delta'~\sigma_\perp~
\right]
\left(
\begin{array}{c}
u^+_k \\
u^-_k
\end{array}
\right)~,
\label{tBdGkLZ}
\ee
where $t'=t+\tau_Q(\cos k-1)/(1+\sin^2k)$ is shifted time, 
$\sigma=(\sigma^y\sin k-\sigma^x)/\sqrt{1+\sin^2k}$ and
$\sigma_\perp=(\sigma^y+\sigma^x\sin k)/\sqrt{1+\sin^2k}$ are
two orthogonal spin components, $\tau=\tau_Q/2\sqrt{1+\sin^2k}$ is the LZ transition time,
and $\Delta'=2\sin k(1-\cos k)/\sqrt{1+\sin^2 k}$ is the minimal gap
at the anticrossing center when $t'=0$. 

In the adiabatic limit, $\tau_Q\gg1$, only the long wavelength modes 
$k\approx0$ get excited with the LZ probability
\be
p_k~=~e^{-\pi\tau(\Delta')^2}~\approx~e^{-\pi\tau_Qk^6/2}~
\label{pkLZ}
\ee
and the density of quasiparticle excitations after crossing the multicritical point is
\be
d~=~\int_{-\pi}^\pi\frac{dk}{2\pi}~p_k~=~
d_0~\tau_Q^{-1/6}~,
\label{d16}
\ee
where $d_0=2^{1/6}\Gamma(7/6)\pi^{-7/6}=0.274$ and $\Gamma(...)$ is the gamma function.

The correct exponent $1/6$ can be made compatible with the general Eq. (\ref{dKZgeneral}) 
as follows. The instantaneous quasiparticle frequency in Eqs. (\ref{tBdGkLZ}) is 
$\omega'_k=\sqrt{(\epsilon')^2+(\Delta')^2}$, where 
$\epsilon'=t'/\tau$ is the relevant distance from the anticrossing center. We can expand 
$\omega'_k\sim|k|^3\equiv|k|^{z'}$ 
at $\epsilon'=0$ and small $k$,  
$\omega'_0\sim|\epsilon'|\equiv|\epsilon'|^{z'\nu'}$ at small $\epsilon'$, to identify the 
exponents relevant for a homogeneous transition as 
\be
z'=3~,~~\nu'=1/3~.
\label{z'nu'}
\ee
Using these relevant exponents in the general Eq. (\ref{dKZgeneral}) gives the correct 
exponent $1/6$ in the exact Eq. (\ref{d16}). Note that both $z$ in Eq. (\ref{znu}) and 
$z'$ in Eq. (\ref{z'nu'}) are greater than $1$.
 
Moreover, when the transition is in space, then 
Eq. (\ref{tildeDeltageneral}) with the exponents (\ref{z'nu'}) predicts the gap
\be
\hat \Delta_{SP} ~\simeq~ \alpha^{3/4}~,
\label{hatDeltaprime}
\ee
while the ``canonical'' exponents (\ref{znu}) give 
$\hat\Delta_{SP}\simeq\alpha^{2/3}$. In Section \ref{XYinspace} we will see that 
Eq. (\ref{hatDeltaprime}) is indeed the correct gap in a transition in space. 

However, depending on the choice of $\nu$ or $\nu'$ in Eq. (\ref{tildexigeneral}) 
we obtain either $\hat\xi_{SP}\simeq\alpha^{1/3}$ or $\hat\xi_{SP}\simeq\alpha^{1/4}$ of 
which only the former will turn out to be relevant for a transition in space, but 
even it is not a unique scale of length there. Nevertheless, since both $z>1$ and 
$z'>1$, we expect the adiabatic threshold velocity $\hat v$ in an inhomogeneous 
transition to scale with a positive power of the inhomogeneity $\alpha$,
compare the general Eq. (\ref{tildevgeneral}) .

\subsection{ Adiabatic regime of homogeneous transition } 
\label{XYhomoadiab}

Since a finite chain of $N$ spins has finite energy gap at the critical
$\epsilon=0$, the homogeneous transition becomes adiabatic above a finite 
$\tau_Q$ when the scaling relation (\ref{d16}) crosses over to exponential decay. 
  
Indeed, in a periodic chain the quasimomenta are quantized as 
$k=\pm\frac{\pi}{N},\pm\frac{3\pi}{N},...$ to satisfy anti-periodic boundary 
conditions for the Jordan-Wigner fermions in the subspace of even parity. 
The $p_k=\exp\left(-\pi\tau_Qk^6/2\right)$ in Eq. (\ref{pkLZ}) 
is a probability to excite a pair of quasiparticles with momenta $(k,-k)$. 
When $\tau_Q$ is large enough, then only the longest 
wavelength pair $\left(\frac{\pi}{N},-\frac{\pi}{N}\right)$ has non-negligible excitation 
probability $p_{\pi/N}=\exp\left(-\pi^7\tau_Q/2N^6\right)$, but even this probability 
becomes exponentially small when $\tau_Q$ is deep enough in the adiabatic regime:
\be
\tau_Q~\gg~\frac{2}{\pi^7}~N^6~.
\label{tauQadiab}
\ee
The transition time required for a homogeneous transition to become
adiabatic grows with the sixth power of the number of spins.   

The adiabatic condition (\ref{tauQadiab}) can be also obtained in a more intuitive 
way from Eq. (\ref{hatxi}), compare similar argument in Ref. \cite{KZIsing}. Indeed, 
average distance between excitations scales as $\hat\xi_{KZ}\sim\tau_Q^{1/6}$ and, 
consequently, the $\tau_Q$ required for the whole chain of $N$ spins to remain defect-
free scales as $N^6$. This simple argument applies to an open chain as well. 

In Ref. \cite{nonlinear} a non-linear generalization $\epsilon(t)={\rm sign}(-t)~|t/\tau_Q|^r$ 
of the linear quench (\ref{tauQ}) was proposed to improve adiabaticity.
Indeed, a similar argument as in Section \ref{Intro} yields 
$d_r\sim\tau_Q^{-\frac{r\nu'}{1+rz'\nu'}}$ as a generalization of 
Eq. (\ref{dKZgeneral}). When $r\gg1/z'\nu'=1$ then 
$d_r\sim\tau_Q^{-1/z'}=\tau_Q^{-1/3}$ is much less than $d_1\sim\tau_Q^{-1/6}$
after a linear quench for long enough $\tau_Q$. In a finite chain the non-linear 
quench becomes adiabatic when   
\be
\tau_Q~\gg~N^3~
\ee
but this is still qubic in $N$.

\section{ Transition in space in XY model } 
\label{XYinspace}

\begin{figure}
\includegraphics[width=0.99\columnwidth,clip=true]{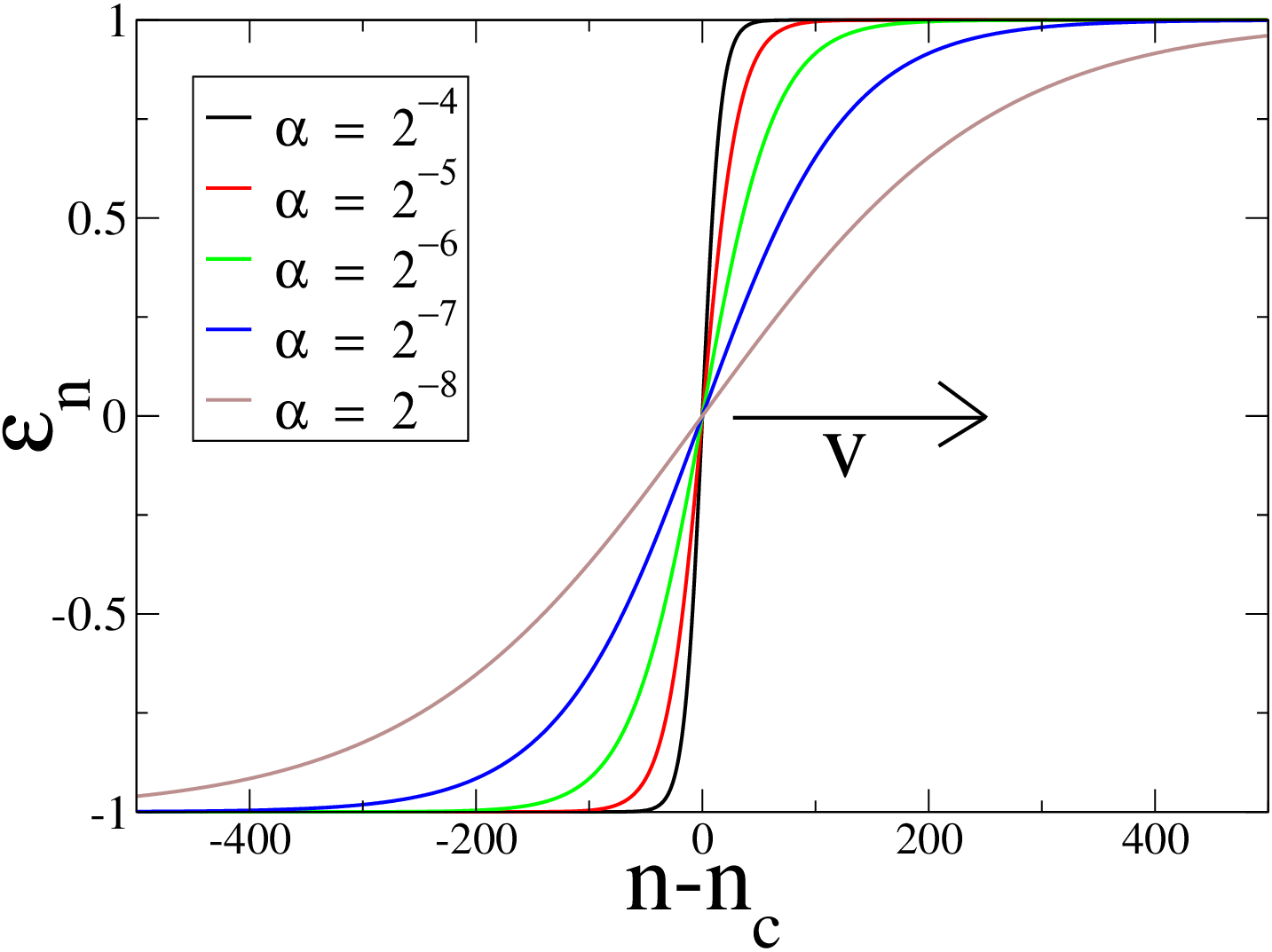}
\caption{ 
The critical front in Eqs. (\ref{slant}) and (\ref{slantv}).
}
\label{Figslant}
\end{figure}

\begin{figure}
\includegraphics[width=0.99\columnwidth,clip=true]{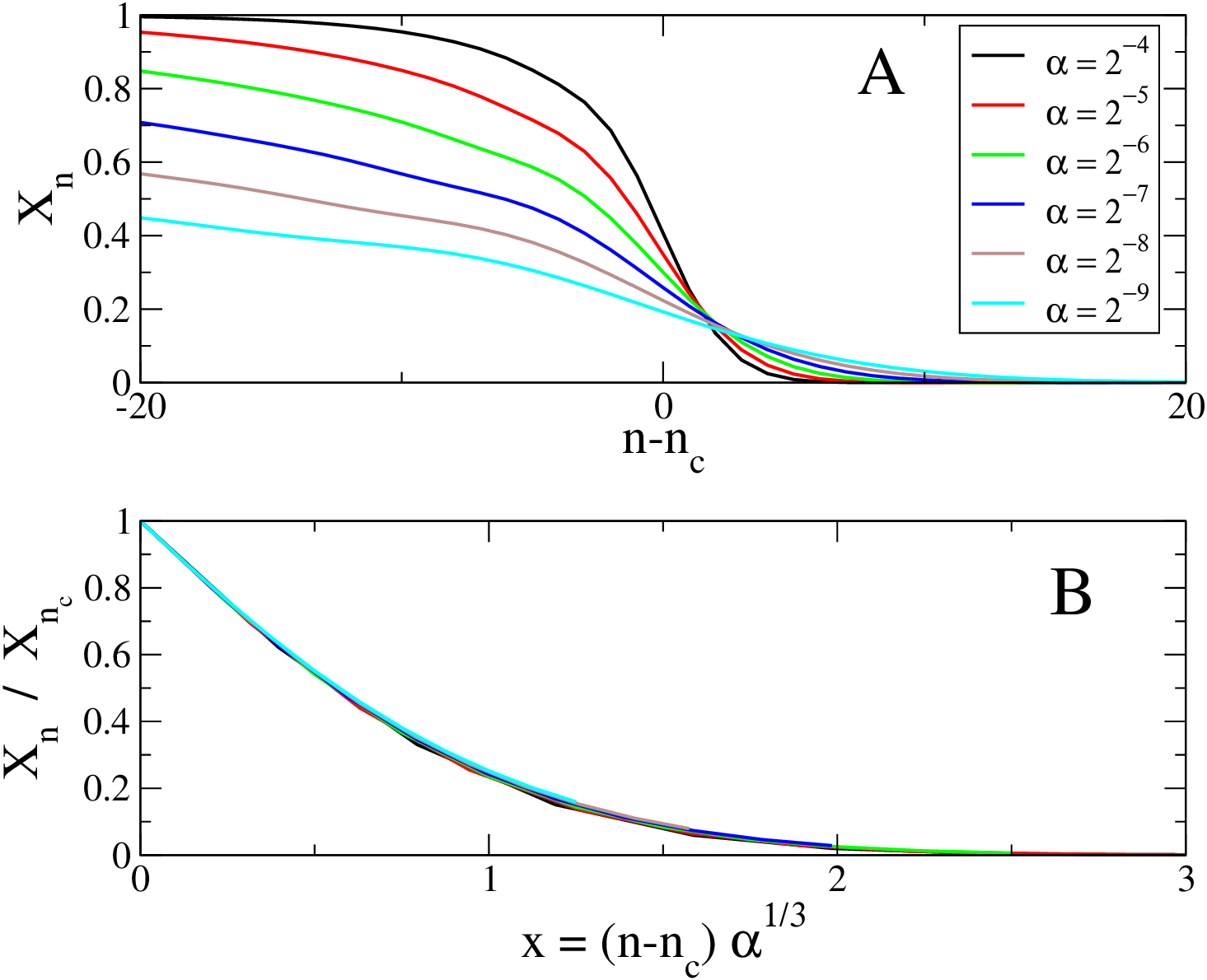}
\caption{ 
Phase transition in space. In panel A, ferromagnetic magnetization 
$X_n=\langle\sigma^x_n\rangle$ is shown as a function of distance from the critical 
point $n-n_c$ for different values of the gradient $\alpha$. This ferromagnetic 
magnetization penetrates into the paramagnetic phase where $n-n_c>0$. In panel B, 
the same magnetization as in panel A but rescaled by its value at the critical point 
$X_{n_c}$, is shown as a function of a rescaled distance $x=(n-n_c)\alpha^{1/3}$. The 
six rescaled plots collapse demonstrating that the magnetization penetrates into the 
paramagnetic phase to a depth $\delta n\simeq\alpha^{-1/3}$ predicted in Eq. (\ref{deltan}). 
}
\label{FIGwnikanie}
\end{figure}

In this Section we take the size of an open chain $N\to\infty$ to avoid boundary 
effects and consider a smooth static slant 
\be
\epsilon_n~=~\tanh\left[\alpha(n-n_c)\right]~\approx~\alpha(n-n_c)~,
\label{slant}
\ee
interpolating between $\epsilon=1$ and $\epsilon=-1$, which is shown in 
Fig. \ref{Figslant}. The slant (\ref{slant}) can be linearized near the 
critical point $n_c$ as in the general Eq. (\ref{nc}). 

We are interested in the low frequency part of the quasiparticle spectrum where, 
presumably, we can make a long wavelength approximation and treat $n$ as a 
continuous variable:
\be
u_{n\pm1}\approx u_n \pm \frac{d}{dn} u_n +\frac12 \frac{d^2}{dn^2} u_n~. 
\label{longwave}
\ee 
We also expect that the low frequency quasiparticle modes are localized near the 
critical point $n_c$, where we can use the linearization in Eq. (\ref{slant}). 

Indeed, after rescaling 
\be
n-n_c=\alpha^{-1/3}x~,~~
\omega=\alpha^{3/4}\Omega~
\label{x}
\ee
equations (\ref{BdG}) become
\bea
&&
\alpha^{\frac{1}{12}}\Omega
\left(
\begin{array}{c}
u^+\\
u^-
\end{array}
\right)-
\sigma^x
\left[
-\frac{d^2}{dx^2}+2x
\right]
\left(
\begin{array}{c}
u^+ \\
u^-
\end{array}
\right)~=
\nonumber\\
&&
\alpha^{1/3}~i\sigma^y
\left[
1+2x\frac{d}{dx}
\right]
\left(
\begin{array}{c}
u^+ \\
u^-
\end{array}
\right).
\eea
To leading order in $\alpha\ll1$ their two eigenmodes of lowest frequency are
\bea
\Omega_0&=&0~,
\label{Omega0}
\\
u^+_0 & \propto & e^{-\frac12\alpha^{1/3}x^2}A(x)~,~~
u^-_0   \propto   0~,
\nonumber
\eea
and 
\bea
\Omega_1 &=& \sqrt{8\Gamma(5/4)/\Gamma(3/4)} ~=~ 2.43~,
\label{Omega1}
\\
u^+_1 & \propto &
e^{-\frac12\alpha^{1/3}x^2}
\frac{dA}{dx}(x)~,~~
u^-_1 \propto
e^{-\frac12\alpha^{1/3}x^2}
\frac{-2A(x)}{\alpha^{\frac{1}{12}}\Omega_1}~,
\nonumber
\eea
where $A(x)={\rm Ai}[2^{1/3}x]$ and Ai is the (oscillating) Airy function. 

The lowest energy relevant (even parity) excitation of the two quasiparticles $\gamma_0$ 
and $\gamma_1$ has a finite gap
\be
\hat\Delta_{SP}~=~
\omega_0+\omega_1 ~=~ 
\alpha^{3/4}\Omega_1 ~\simeq~
\alpha^{3/4}~,
\label{HatDeltaBulk}
\ee
in agreement with the general Eq. (\ref{hatDeltaprime}) for the choice $z'=1/\nu'=3$
in Eq. (\ref{z'nu'}).

However, the modes (\ref{Omega0},\ref{Omega1}) do not have a unique scale of length.
They penetrate into the paramagnetic phase, where $n>n_c$ and $x>0$, to a depth 
\be
\delta n~\simeq~\alpha^{-1/3}~
\label{deltan}
\ee 
determined by the $x\to+\infty$ asymptote of the Airy function 
$A(x)\sim\exp(-2\sqrt2 x^{3/2}/3)$ and Eq. (\ref{x}). Consequently, this $\delta n$ is 
also penetration depth of ferromagnetic magnetization into the paramagnetic phase, see 
Fig. \ref{FIGwnikanie}. However, on the ferromagnetic side, where $n<n_c$ and $x<0$, the 
same modes (\ref{Omega0},\ref{Omega1}) extend to the depth
\be
\Delta n ~\simeq~ \alpha^{-1/2}
\label{Deltan}
\ee
limited by the Gaussian envelope 
$e^{-\frac12\alpha^{1/3}x^2}=e^{-\frac12\alpha(n-n_c)^2}$. This envelope is damping 
oscillations of the Airy function $A(x)$ which take place on the shorter scale 
$\delta n\simeq\alpha^{-1/3}$. An overall width of the modes (\ref{Omega0},\ref{Omega1}) 
is set by the longer scale $\Delta n$. 

Thus in case of a finite chain size $N$ results of this Section require $\Delta n\ll N$ 
or, equivalently, a lower bound $\alpha\gg N^{-2}$.

\section{ Inhomogeneous transition in XY model }
\label{XYinhomo} 

In this Section we make the critical front in Eq. (\ref{slant}) and Fig. \ref{Figslant}
sweep from $n_c\to-\infty$ to $n_c\to+\infty$ at a constant velocity $v$:
\be
\epsilon_n(t)~=~\tanh\left[\alpha(n-vt)\right]~\approx~\alpha(n-vt)~,
\label{slantv}
\ee
see Fig. \ref{Figslant}. Near the critical point at $n=vt$ the slant (\ref{slantv}) 
can be linearized as in the general Eq. (\ref{alpha}).

\subsection{ The homogeneous regime of inhomogeneous transition }
\label{XYinhomohomo} 

We can obtain quasiparticle group velocity at the critical point $\epsilon=0$ from the 
dispersion (\ref{omegak}). The group velocity is maximized for $k = \pm \pi/2$ by
\be
v_q~=~2~.
\ee
When $v\gg v_q$ there is no causal connection across the critical point
and the inhomogeneous transition proceeds as if it were effectively homogeneous with 
a quench time $\tau_Q=1/\alpha v$. In this regime we expect the ``homogeneous'' 
$1/6$-scaling in Eq. (\ref{d16}) to apply and a rescaled final density of kinks
to scale as
\be
\alpha^{-1/6}~ d~\simeq~v^{1/6}~
\label{v16}
\ee
with velocity of the critical front $v$.

\begin{figure}
\includegraphics[width=0.99\columnwidth,clip=true]{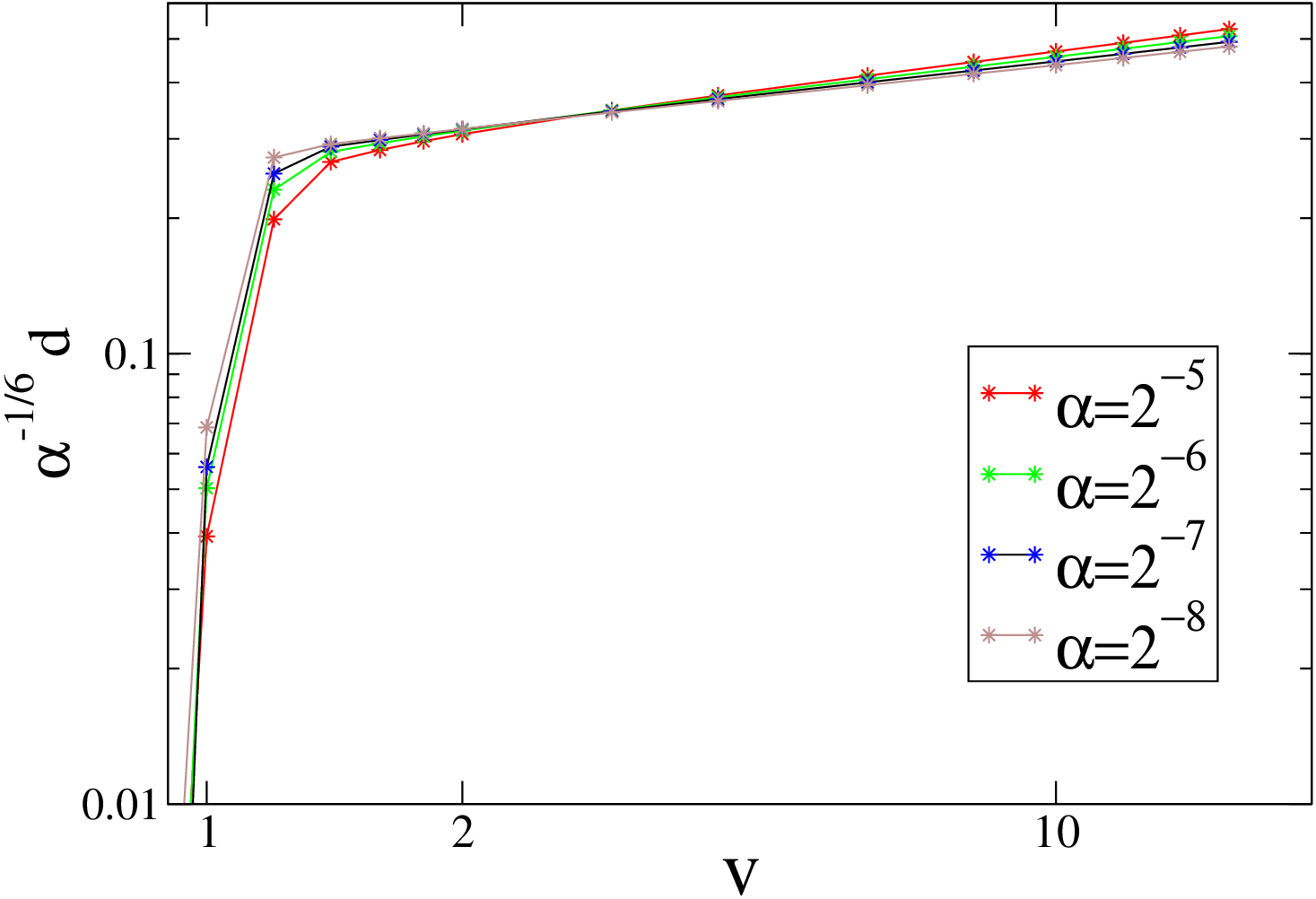}
\caption{ 
Numerical simulations of the time-dependent Bogoliubov-de Gennes equations (\ref{tBdG}) 
in a finite chain of $N=200$ spins. Rescaled final kink density $\alpha^{-1/6}d$ after 
an inhomogeneous transition in Eq. (\ref{slantv}) is shown for different values of the 
gradient $\alpha$. In the homogeneous regime $v\gg2$ the collapsed plots are nearly 
linear with a slope $0.20$ for $\alpha=1/256$ close to the predicted $1/6$ in 
Eq. (\ref{v16}) where $\alpha^{-1/6}d\simeq v^{1/6}$. 
}
\label{Figdv}
\end{figure}

To verify this prediction we simulated time-dependent Bogoliubov-de Gennes equations
\bea
i\frac{d}{dt} u^\pm_n &=&
2(1+\epsilon_n) u^\mp_n 
\nonumber\\
&-&
(1\mp\epsilon_{n+\frac12}) u^\mp_{n+1}-(1\pm\epsilon_{n+\frac12}) u^\mp_{n-1}   
~.
\label{tBdG}
\eea
in an open chain of $N$ spins. Results are shown in Fig. \ref{Figdv} and in the 
homogeneous regime $v\gg2$ they are consistent with the prediction (\ref{v16}).

\subsection{ The adiabatic regime of inhomogeneous transition } 
\label{XYinhomoadiab} 

In this Section we consider the adiabatic limit of small front velocity $v$ when only 
the lowest relevant excited state has non-negligible excitation probability. The state 
is the even parity state occupied by the lowest two quasiparticles: $\gamma_0$ and 
$\gamma_1$. When the critical point $n_c$ is in the bulk of the finite lattice, then 
these quasiparticles are described by the Bogoliubov modes (\ref{Omega0},\ref{Omega1}) 
and the energy gap for this excitation is given by Eq. (\ref{HatDeltaBulk}).

\begin{figure}
\includegraphics[width=0.99\columnwidth,clip=true]{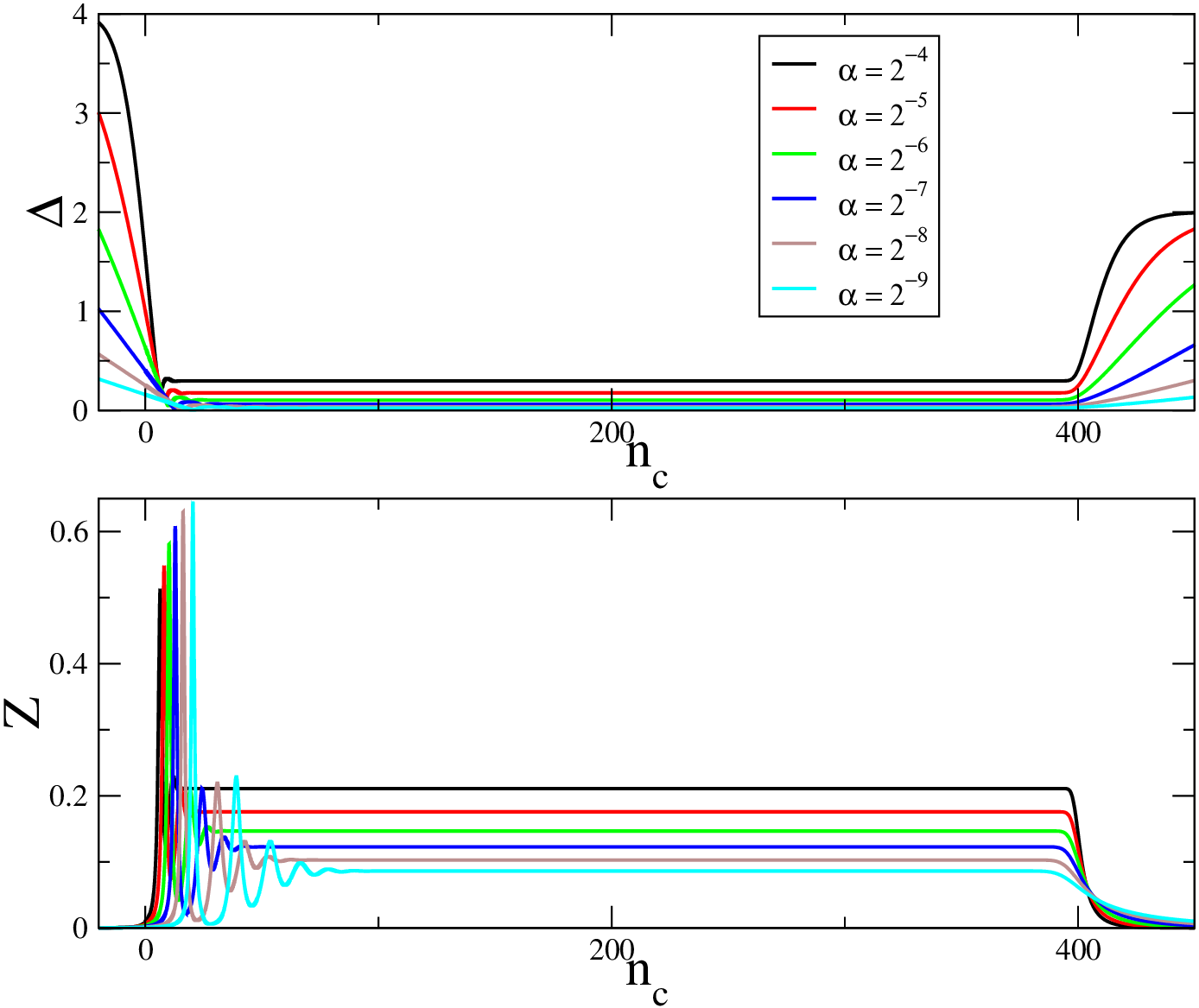}
\caption{ 
Parameters $\Delta$ and $Z$ of the LZ problem in Eq. (\ref{LZ}) for a lattice size $N=400$ 
and different values of the small gradient $\alpha$. In panel A, the instantaneous relevant 
gap $\Delta=\omega_0+\omega_1$ is shown as a function of $n_c$. The bulk value of the gap, 
when $1\ll n_c \ll N$, is estimated in Eq. (\ref{HatDeltaBulk}) as $\hat\Delta_{SP}\simeq\alpha^{3/4}$. 
In panel B, the parameter $Z$ in Eq. (\ref{Z}) is shown. Its bulk value can be estimated from 
Eqs. (\ref{Omega0},\ref{Omega1},\ref{Z}) as $Z\simeq\alpha^{1/4}$. In both panels A and B, 
the crossover regions at both ends of the chain have a width $\Delta n\simeq\alpha^{-1/2}$ 
set by the Gaussian envelope of the modes (\ref{Omega0},\ref{Omega1}). The additional 
oscillations at the left end have a length scale $\delta n\simeq\alpha^{-1/3}$ determined by 
the oscillatory Airy function $A(x)$ in the modes (\ref{Omega0},\ref{Omega1}).  
}
\label{FigDeltaZ}
\end{figure}

In the adiabatic limit the even parity subspace of the Hilbert space can be truncated 
to an effective two-level system:
\be
|\psi(t)\rangle~=~
a(t)~|0\rangle~+~
b(t)~|1\rangle~,
\ee
where $|0\rangle$ is the instantaneous ground state in the even subspace for an
instantaneous position $n_c$ of the critical point and 
$|1\rangle=\gamma_1^\dag\gamma_0^\dag|0\rangle$ is the instantaneous first excited
state for the same $n_c$. The amplitudes $a,b$ solve a generalized Landau-Zener (LZ) 
problem
\bea
i\frac{d}{dt}
\left(
\begin{array}{c}
a\\
b
\end{array}
\right)
~=~
\left(
\begin{array}{cc}
0    & ivZ    \\
-ivZ & \Delta
\end{array}
\right)
\left(
\begin{array}{c}
a\\
b
\end{array}
\right)
\label{LZ}
\eea
with initial conditions $a(-\infty)=1$ and $b(-\infty)=0$. Here $\Delta=\omega_0+\omega_1$ is 
an instantaneous gap and
\bea
Z \equiv
\langle1|\frac{d}{dn_c}|0\rangle =
\sum_{n=1}^N
\left(v_{n1},u_{n1}\right)
\frac{d}{dn_c}
\left(
\begin{array}{c}
u_{n0} \\
v_{n0}
\end{array}
\right)~.
\label{Z}
\eea
Generic $\Delta(n_c)$ and $Z(n_c)$ are shown in Figs. \ref{FigDeltaZ} A and B respectively.

\begin{figure}
\includegraphics[width=1.0\columnwidth,clip=true]{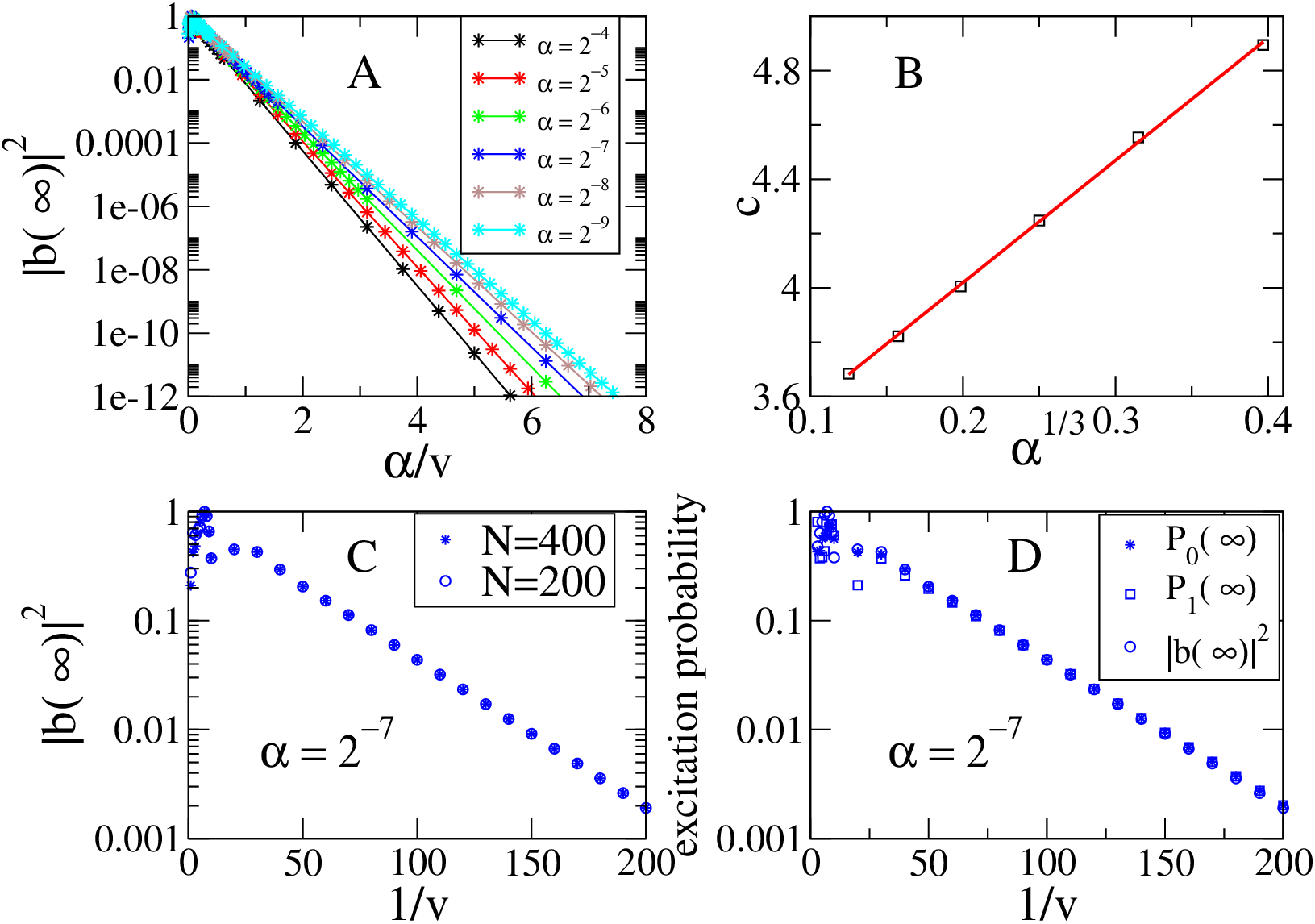}
\caption{ 
In panel A, the final excitation probability $|b(\infty)|^2$ in the generalized LZ 
problem, defined in Eq. (\ref{LZ}) and 
Fig. \ref{FigDeltaZ}, as a function of $\alpha/v$ for different values of the 
inhomogeneity $\alpha$. Here the solid lines are the best fits of the LZ formula 
(\ref{PLZ}) to the numerical data points: 
$|b(\infty)|^2=\exp\left(-c\frac{\alpha}{v}\right)$.
In panel B, the coefficients $c$ fitted in panel A as a function of $\alpha$. The solid 
line is the best fit $c=3.12+4.50\alpha^{1/3}$ demonstrating weak residual dependence 
on $\alpha^{1/3}$ which becomes 
negligible when $\alpha\to 0$ and $c\to3.12$.
In panel C, $|b(\infty)|^2$ as a function of $1/v$ for a fixed 
$\alpha=2^{-7}$ and two different chain sizes $N=200,400$. In the adiabatic regime the 
excitation probability does not depend on $N$ demonstrating that the excitation of the 
lowest two quasiparticles 
is a boundary effect.
In panel D, $|b(\infty)|^2$ from the LZ model and corresponding excitation 
probabilities $P_0=\langle\gamma_0^\dag\gamma_0\rangle$ and  
$P_1=\langle\gamma_1^\dag\gamma_1\rangle$ from the exact Bogoliubov-de Gennes 
equations (\ref{tBdG}). All these three probabilities are the same,
$|b(\infty)|^2=P_0=P_1$, in the adiabatic regime where only a pair of the lowest 
two quasiparticles $\gamma_0$ and $\gamma_1$ can get excited.
}
\label{FigAdiab}
\end{figure}

In the adiabatic limit of small excitation probability $|b(\infty)|^2\ll1$, our numerical 
simulations of the LZ equations (\ref{LZ}) are well described by a simple 
LZ-like formula 
\be
|b(\infty)|^2~=~\exp\left(-c~\frac{\alpha}{v}\right)~,
\label{PLZ}
\ee
where $c=O(1)$ is a numerical prefactor. Indeed, the plots for different $\alpha$ which 
are collected in Fig. \ref{FigAdiab} A nearly collapse. The collapse is not 
perfect because, as shown in Fig. \ref{FigAdiab} B, there is still a weak residual 
dependence $c\approx3.12+4.50~\alpha^{1/3}$. However, when $\alpha\ll1$ then $c\approx3.12$ 
becomes independent of $\alpha$ as assumed in Eq. (\ref{PLZ}). 

Furthermore, Fig. \ref{FigAdiab} C shows that in the adiabatic regime the small 
excitation probability $|b(\infty)|^2$ does not depend on the chain size $N$. Not 
quite surprisingly, the excitation of the lowest two quasiparticles is a boundary 
effect determined by the behavior of $\Delta$ and $Z$ in Fig. \ref{FigDeltaZ} when 
the critical point $n_c$ is near the ends of the chain. 

We can conclude that the excitation probability (\ref{PLZ}) is exponentially small 
when $v\ll\alpha$. This inequality identifies a threshold velocity 
\be
\hat v ~\simeq~ \alpha
\ee 
when the inhomogeneous transition becomes adiabatic. As anticipated for $z>1$, the 
adiabatic threshold $\hat v$ is a positive power of the gradient $\alpha$.

\subsection{ Residual edge excitations after adiabatic transition } 
\label{XYinhomoedge} 

In the adiabatic limit $v\ll\alpha$ only the lowest two quasiparticles 
$\gamma_0$ and $\gamma_1$ can get excited. Since the inhomogeneous 
transition is between two gapped phases, these low frequency modes 
are localized either near the critical point $n_c$ when $n_c$ is in 
the bulk of the spin chain, or at one of the ends of the chain when 
$n_c$ is near this end. For instance, the ``bulk'' modes 
in Eqs. (\ref{Omega0},\ref{Omega1}) are localized within the distance 
$\Delta n\simeq\alpha^{-1/2}$ from the critical point.

\begin{figure}
\includegraphics[width=0.99\columnwidth,clip=true]{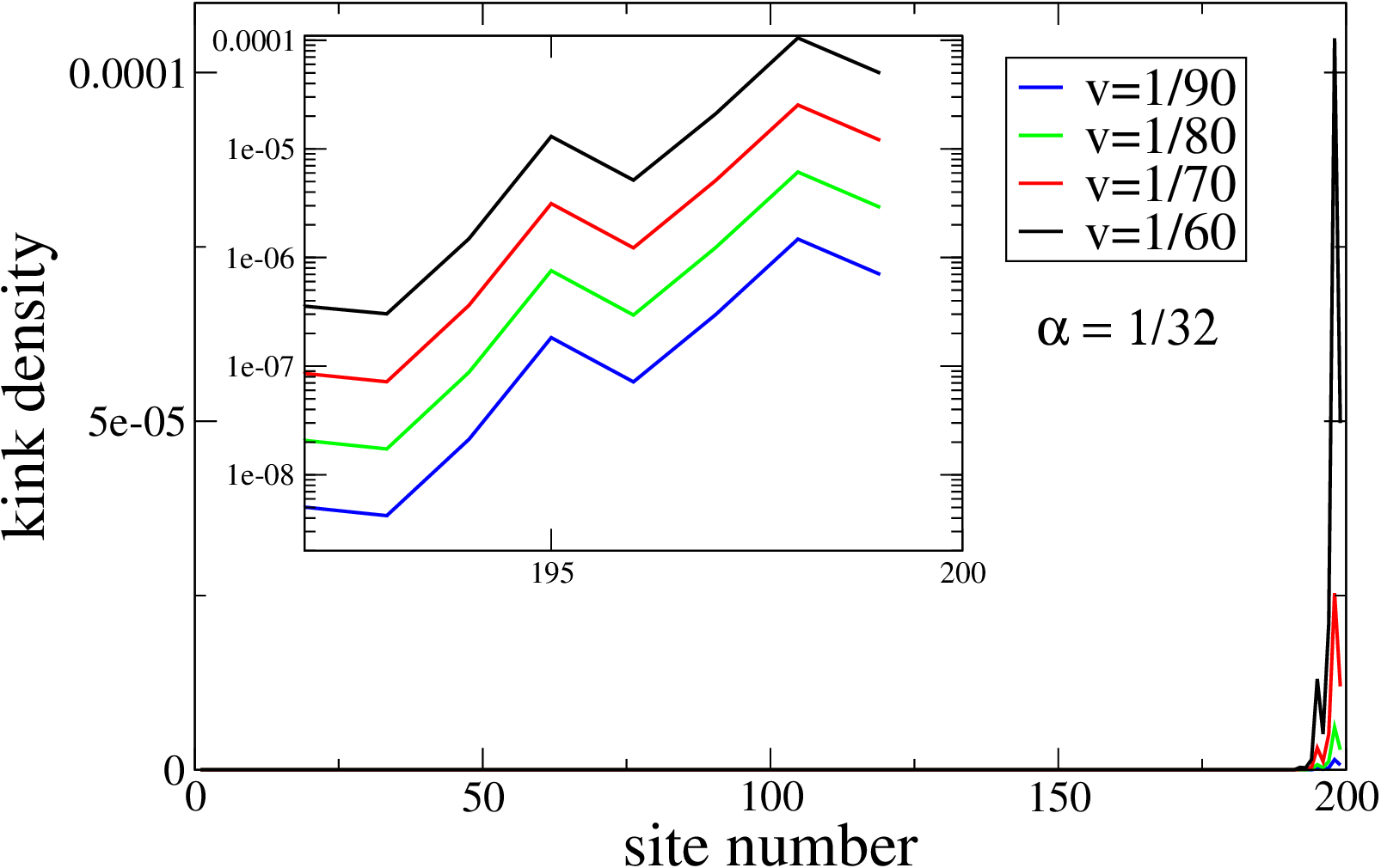}
\caption{ 
Density distributions of kinks along a chain of $N=200$ spins after an 
inhomogeneous transition in the adiabatic regime $v\ll\alpha$. 
These distributions were obtained from numerical simulations of the exact 
time-dependent Bogoliubov-de Gennes equations (\ref{tBdG}) with the critical 
front in Fig. \ref{Figslant}. Residual (exponentially small) excitations 
are brushed away by the critical front to the right edge leaving behind 
defect-free bulk of the chain.
The inset is a log-scale focus on the right edge. It shows that in the adiabatic 
regime all density distributions for different $v$ are the same up to 
an overall $v$-dependent amplitude set by the Landau-Zener excitation probability 
in Eq. (\ref{PLZ}).  
}
\label{FigEdge}
\end{figure}

Consequently, as the critical front in Fig. \ref{Figslant} is passing 
across the chain these instantaneous low frequency modes follow the moving 
front to the right end of the chain. Indeed, a few generic final density distributions 
of kinks along the spin chain are shown in Fig. \ref{FigEdge}. These (exponentially 
small) probabilities of kink excitation are localized near the right end of the 
chain. Residual excitations, if any, are brushed away to the right end 
leaving behind a defect-free bulk of the chain.

\section{ Conclusion } 
\label{Conclusion}

Let us itemize what we know about dynamics of a quantum phase transitions across the 
multicritical point of the XY chain which is an example of a transition with $z>1$:

\bi

\item Density of excitations after a homogeneous transition in an infinite chain decays 
with the sixth root of the transition time $\tau_Q$. 

\item Consequently, the transition time required for a homogeneous transition in 
a finite chain of $N$ spins to be adiabatic scales like $N^6$.

\item Quasiparticles excited by a homogeneous transition are uniformly scattered 
along a spin chain.

\item In an inhomogeneous transition, when the critical front propagates much faster than 
the maximal quasiparticle group velocity $v_q=2$, then the transition is effectively homogeneous.

\item When a critical front moves much slower than the adiabatic threshold velocity 
$\hat v\simeq\alpha$, then average number of excitations is exponentially suppressed 
and does not depend on $N$. 

\item Quasiparticles excited in the adiabatic regime, if any, are brushed away by 
the critical front to the end of a spin chain leaving behind defect-free
bulk of the chain. 

\item The minimal time required for the adiabatic inhomogeneous transition to be 
completed is $N/\hat v$. It is a mere linear function of $N$ instead of the 
$N^6$ in the homogeneous case.

\ei

The main difference between the case $z=1$ in Ref. \cite{inhomRams} and $z>1$ considered
here is that in the former case the adiabatic threshold velocity $\hat v$ is the same as 
the homogeneous threshold velocity $v_q$, $\hat v=v_q$ when $z=1$, while in the latter
case the adiabatic threshold is less than the homogeneous one, $\hat v<v_q$ when $z>1$,
and it is proportional to the inhomogeneity of the transition. When $z>1$ the KZ adiabatic
threshold $\hat v$ is nonzero despite vanishing minimal quasiparticle group velocity.  

Putting the results for $z>1$ together with those of Ref. \cite{inhomRams} for $z=1$, we can 
conclude that for large $N$ an inhomogeneous transition is a more efficient method 
of adiabatic quantum state preparation than a straightforward homogeneous transition. 
Not only the time required for an adiabatic transition is much shorter, but also any 
residual (exponentially small) excitations are brushed away to the end of the spin 
chain.

\section*{ Acknowledgements } 

We thank Wojciech Zurek for discussions. Work of J.D. and M.M.R. was supported in part
by the Polish Ministry of Science and Education research projects N202 079135 and 
N202 174335 respectively. M.M.R. acknowledges partial support of the U.S. Department of 
Energy through the LANL/LDRD Program


\end{document}